\documentclass[preprint,aps,pre,showpacs,superscriptaddress]{revtex4}

\def \beq{\begin{equation}}         \def \eeq{\end{equation}}
\def \beqa{\begin{eqnarray}}        \def \eeqa{\end{eqnarray}}
\def \bea{\begin{array}}        \def \eea{\end{array}}

\def\bio#1#2#3{{Biophys. J. }{\bf #1}, #2 (#3)}

\def\prl#1#2#3{{Phys. Rev. Lett. }{\bf #1}, #2 (#3)}
\def\sci#1#2#3{{Science }{\bf #1}, #2 (#3)}

\usepackage{amsmath}
\usepackage{epsfig}
\begin{document}

\title{Bayesian analysis of time series of single
RNA under fluctuating force}
\author{Fei Liu}
\email[Email address:]{liufei@tsinghua.edu.cn} \affiliation{Center
for Advanced Study, Tsinghua University, Beijing, 100084, China}
\author{Zhong-can Ou-Yang}
\affiliation{Center for Advanced Study, Tsinghua University,
Beijing, 100084, China} \affiliation{Institute of Theoretical
Physics, The Chinese Academy of Sciences, P.O.Box 2735 Beijing
100080, China}
\date{\today}

\begin{abstract}
Extracting the intrinsic kinetic information of biological
molecule from its single-molecule kinetic data is of considerable
biophysical interest. In this work, we theoretically investigate
the feasibility of inferring single RNA's intrinsic kinetic
parameters from the time series obtained by forced
folding/unfolding experiment done in the light tweezer, where the
molecule is flanked by long double-stranded DNA/RNA handles and
tethered between two big beads. We first construct a coarse-grain
physical model of the experimental system. The model has captured
the major physical factors: the Brownian motion of the bead, the
molecular structural transition, and the elasticity of the handles
and RNA. Then based on an analytic solution of the model, a
Bayesian method using Monte Carlo Markov Chain is proposed to
infer the intrinsic kinetic parameters of the RNA from the noisy
time series of the distance or force. Because the force
fluctuation induced by the Brownian motion of the bead and the
structural transition can significantly modulate the transition
rates of the RNA, we prove that, this statistic method is more
accurate and efficient than the conventional histogram fitting
method in inferring the molecule's intrinsic parameters.
\end{abstract}

\pacs{87.15.Aa, 82.37.Rs, 87.15.By, 82.20.Uv} \maketitle

The current Single-molecule manipulation provides a novel approach
to study the kinetics of single RNA. Different from many
conventional experimental techniques, such as X-ray
crystallograph, which usually only provide static pictures of the
molecule, the current manipulation techniques, mainly including
the optical tweezer, can trace the full folding/unfolding
processes of single RNA by monitoring the molecule's extension or
force exerted on it in real time~\cite{Liphardt01,Woodside,Wen}.

\begin{figure}[htpb]
\begin{center}
\includegraphics[width=0.9\columnwidth]{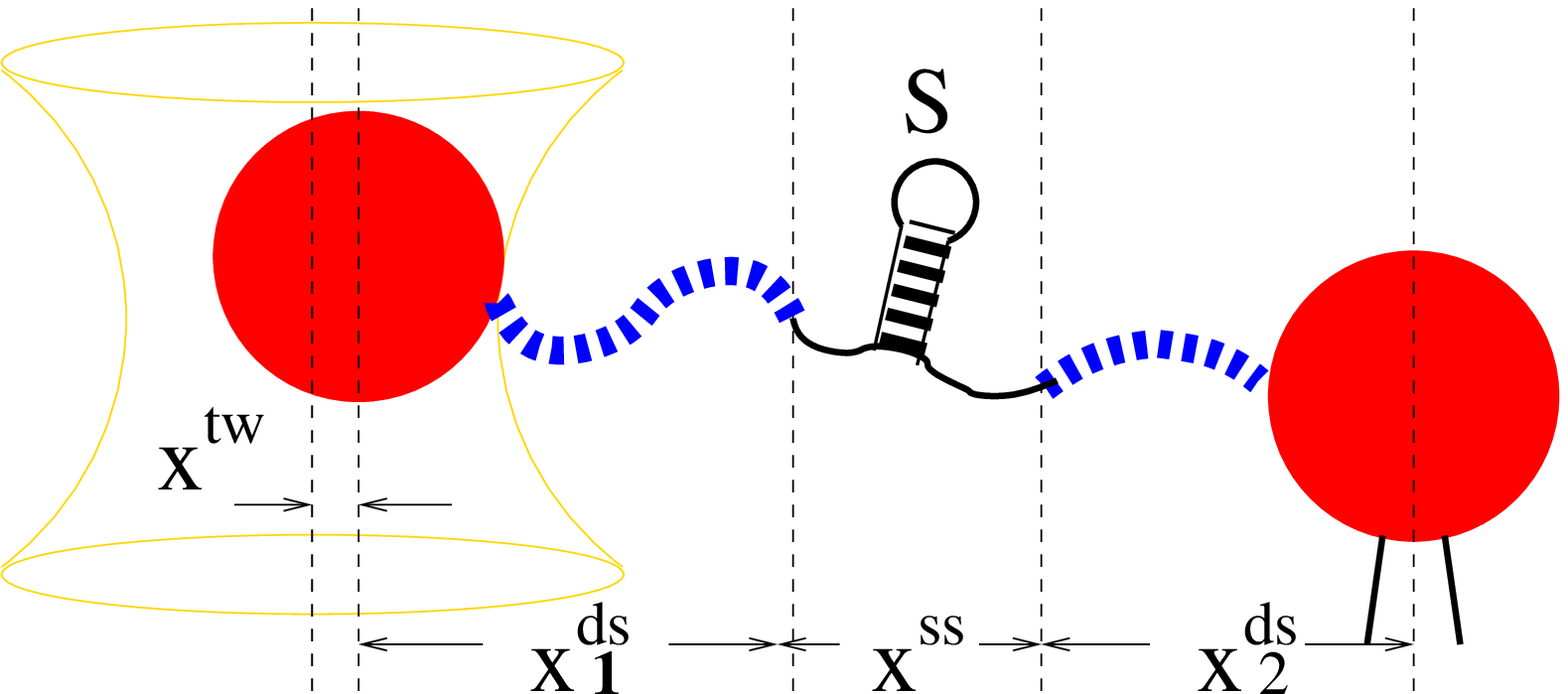}
\caption{} \label{figure1}
\end{center}
\end{figure}

As many nano- or mesoscopic systems, the behavior of single RNA
($\sim$30 nm) in light tweezer is highly dynamic and noisy. The
situation could become more complicated in practice: in order to
manipulate single RNA by the optical trapping method, the RNA must
first be tethered between two large dielectric beads
($\sim$$\mu$m) through two long double-stranded DNA/RNA handles
($\sim$$\mu$m); see Fig.~\ref{figure1}. Due to the presence of the
beads and handles, it would be expected that the kinetics of the
RNA observed in the light tweezer experiment is distinct from the
kinetics of the linker-free RNA. Hence, how to extract the
intrinsic kinetic information of single RNA from experimental data
is an intriguing biophysical issue. One of the possible strategies
is to find optimal experimental conditions through experimental
comparison and computational simulation~\cite{Wen,Manosas07}.
Alternative way is to collect the existing RNA kinetic data and
infer the intrinsic parameters by advanced statistic approaches.
To the best of our knowledge, the latter was not quantitatively
implemented in literature. In this Communication, we present such
an effort.\\

{\noindent\bf Physical model.} Forced folding/unfolding single
RNAs could be achieved in two types of manipulation experiments.
One is the constant force mode (CFM), where the experimental
control parameter, a constant force $F$ of preset value, is
applied on the bead in the light tweezer with or without feedback
control~\cite{Wen,Woodside}. The other is the passive mode (PM),
where the control parameter, the distance between the centers of
the light tweezer and the bead held by the micropipette, $x_{\rm
T}$, is left stationary (see Fig.~\ref{figure1}). The RNA and
light tweezer system involves several time scales: the relaxation
time of the bead in the tweezer, $\tau_{\rm b}$, the relaxation
time of the handles and single-stranded (ss) RNA, $\tau_{\rm h}$
and $\tau_{\rm ssRNA}$, the characteristic time of the overall
kinetics of the RNA, $\tau_{\rm f-u}$, and the characteristic time
of the opening/closing of single base pairs $\tau_{\rm
bp}$~\cite{Manosas05,Manosas07}. Under the conventional
experimental conditions~\cite{Liphardt01,Wen,Woodside}, the
relaxation time $\tau_{\rm h}$, $\tau_{\rm ssRNA}$ and $\tau_{\rm
bp}$ is always far shorter than the relaxation time of the bead
and overall RNA kinetics~\cite{Manosas05,Manosas07}. It is
plausible to assume that the RNA is two-state, i.e., folded (f) or
unfolded (u), and the extension of the handles and ssRNA is in
thermal equilibrium instantaneously. Note that we do not require
that the relaxation of the bead in the light tweezer is also
instantaneous.

Our model involves two freedom degrees: one is the state of the
RNA; the other is the distance $x$ between the centers of the two
beads. Because the force directly controlling the kinetics of the
RNA is always fluctuating with time, we describe the experimental
system by the following two coupled diffusion-reaction equations:
\begin{eqnarray}
\label{diffusionreactonequations}
&& \frac{\partial}{\partial
t}P_{\rm f}(x,t)=\left[{\cal L}_{\rm f}-k^{\rm u}(x)\right]P_{\rm f}+
k^{\rm f}(x)P_{\rm u},\\
&&\frac{\partial}{\partial t}P_{\rm u}(x,t)=\left[{\cal L}_{\rm
u}-k^{\rm f}(x)\right]P_{\rm u}+ k^{\rm u}(x)P_{\rm f},\nonumber
\end{eqnarray}
where $P_{\rm i}(x,t)$ is the probability distribution of the RNA
at state i (f or u) and the distance having a particular value $x$
at time $t$. The Fokker-Planck operators ${\cal L}_{\rm i}$ in the
above equations are
\begin{eqnarray}
\label{definitionFKoperator} {\cal L}_{\rm i}=D
\frac{\partial}{\partial x}e^{-\beta V_{\rm
i}(x)}\frac{\partial}{\partial x}e^{\beta V_{\rm i}(x)},
\end{eqnarray}
where $D$ is diffusion coefficient, $\beta^{-1}=k_{\rm B}T$ with
$k_{\rm B}$ being the Boltzmann's constant and $T$ the absolute
temperature; $V_{\rm i}(x) $ is the RNA state-dependent potential
and defined as $V_{\rm i}(x)=W_{\rm ext} (x) + \int_0^x {f_{\rm i}
(x')dx'}$ with $f_{\rm i} (x) =  \left[ 0.25 \left ( 1 - x/l_{\rm
i} \right)^{-2} + x/l_{\rm i} - 0.25\right]\left/\beta P_{\rm
eff}^{\rm i}\right.$~\cite{Marko,Bustamante} with the persistent
length $P_{\rm eff}^{\rm i}$~\cite{liuf1} and contour length
$l_{\rm i}=2L_{\rm h}+L_{\rm ssRNA}^{\rm i}$; and the external
work $W_{\rm ext} (x)$ done by the external force is $Fx$ in the
CFM and $ \varepsilon (x_{\rm T}-x)^2\left/2\right.$ with a
tweezer stiffness $\varepsilon$ in the PM, respectively. For the
``reaction'' rates $k^{\rm i}(x)$, though there are significant
debates about the correctness of the Bell formula,
$k(f)=k_0\exp[\beta fx^{\ddag}]$~\cite{Bell} in describing
biological molecule's rupture or unfolding, where $k_0$ is the
intrinsic rate constant in the absence of force, and $x^{\ddag}$
is the transition state location, we still use this
phenomenological formula with a slight modification rather than
other improved rate models having certain microscopic
explanation~\cite{Dembo,Evans97,Shapiro,Dudkoprl}. Our
consideration is as follows. First the Bell formula is still the
simplest and most widely used in single molecule studies.
Particularly, it seems to work quite well in the real RNA
folding/unfodling experiments~\cite{Liphardt01,Woodside,Wen}.
Second, other rate formulas are all model-dependent; whether they
are indeed suitable to the ``macroscopic" RNA folding/unfolding is
not undoubted. The rate invoked here is $k^{\rm u}(x) =  k^{\rm
u}_0\exp\left[\beta f_{\rm f}(x)d^\ddag_{\rm f}\right]$ for
$k^{\rm u}\le k_{\rm max }$, otherwise $k^{\rm u}(x)=k_{\rm max}$,
where $k^{\rm u}_0$ and $d^\ddag_{\rm f}$ are respectively the
intrinsic unfolding rate in the absence of force and the
transition state location away from the folded RNA state. This
modification is necessary, in that the unfolding rate given by the
Bell formula increases too fast with force~\cite{liuf2}.
Interestingly, it is not a problem for the folding rate, $k^{\rm
f}(x)=k^{\rm f}_0\exp\left[-\beta f_{\rm u}(x)d^\ddag_{\rm
u}\right]$, and $k^{\rm f}_0$ and $d^\ddag_{\rm u}$ are the
intrinsic folding rate in the absence of force and the transition
state location away from the unfolded RNA state, respectively.

Eq.~\ref{diffusionreactonequations} has an exact solution under
the steady-state assumption of the system:
\begin{eqnarray}
P^{\rm ss}_{\rm i}(x)=\pi_{\rm i} p^{\rm eq}_{\rm i}(x),
\end{eqnarray}
where
\begin{eqnarray}
p^{\rm eq}_{\rm i}(x)= \exp[-\beta V_{\rm i}(x)]\left/\int
\exp[-\beta V_{\rm i}(x')]dx' \right.,
\end{eqnarray}
$\pi_{\rm i}=\langle k^{\rm i}\rangle_{\bar{\rm i}}\left/\langle
k\rangle\right.$, $\bar {\rm i}={\rm f,u}$ respectively correspond
to $\rm i=u,f$, the symbol $\langle\rangle_{\rm i}$ is the average
over the distribution $p^{\rm eq}_{\rm i}(x)$, and $\langle
k\rangle =\langle k^{\rm u}\rangle_{\rm f}+\langle k^{\rm
f}\rangle_{\rm u}$. Obviously, ${\cal L}_{\rm i}p^{\rm eq}_{\rm
i}(x)=0$. Because the experiments are usually carried out under
the steady-state condition, these definition and formulas would be
useful in deeply understanding the RNA forced folding/unfolding
kinetics.

\begin{figure}[htpb]
\begin{center}
\includegraphics[width=0.9\columnwidth]{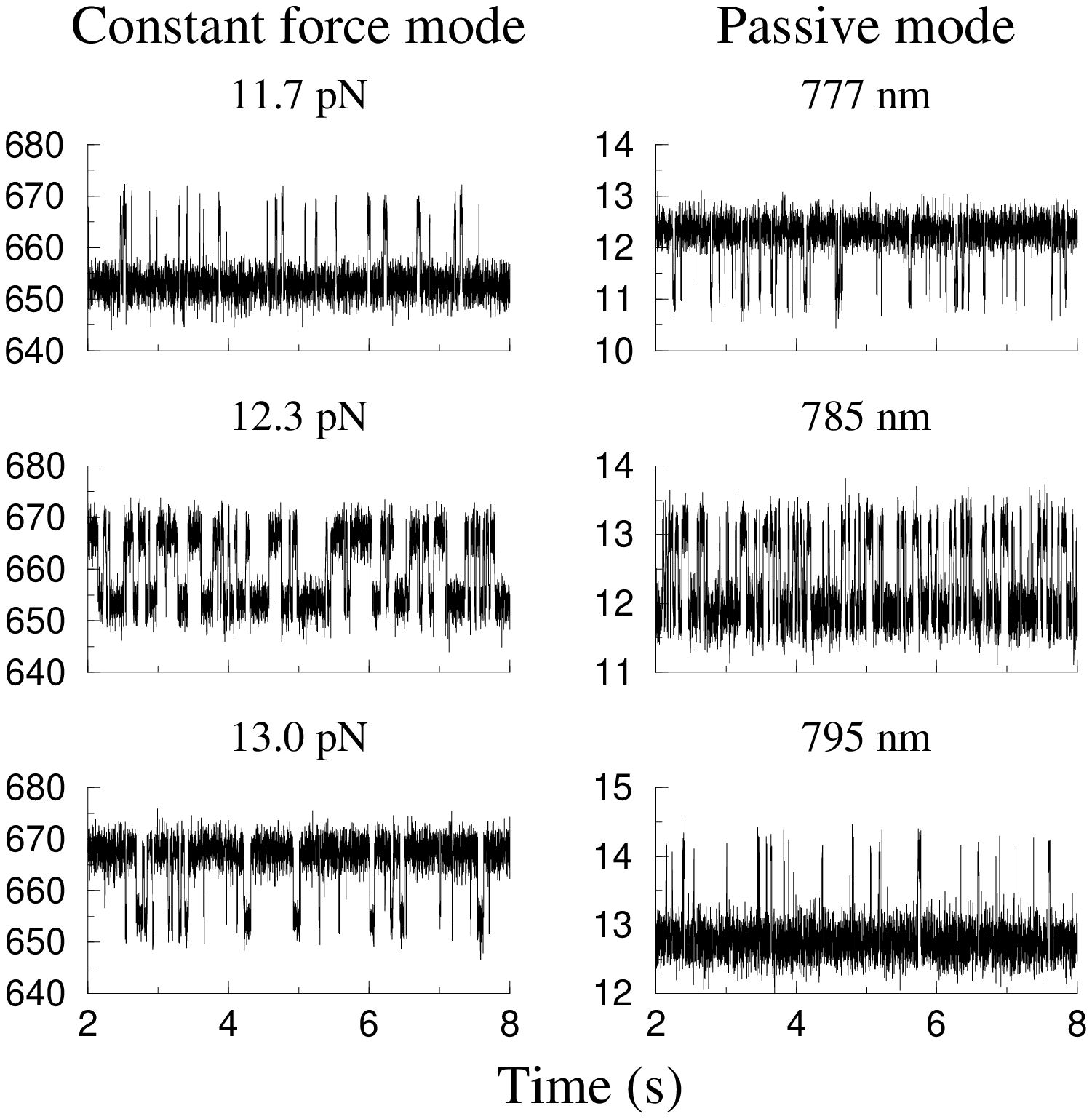}
\end{center}
\caption{}\label{figure2}
\end{figure}

In general, Eq.~\ref{diffusionreactonequations} does not have
exact time-dependent solutions except the rapid diffusion limiting
discussed below~\cite{liuf3}. We have to seek simulation approach
for general situations. Fig.~\ref{figure2} shows several time
series of the distance $x$ or the force $f$ exerted by the tweezer
in the CFM and PM, respectively, and the time interval is 1 ms.
The simulation parameters used are $\varepsilon=0.1$ pN/nm for the
tweezer stiffness, $R_{\rm b}=1.0$ $\mu$m for the bead radius;
$\eta=10^{-3}$ kg/ms for the viscosity of water, $L_{\rm h}=340.0$
nm (1000 base-pairs) and $P_{\rm h}=53.0$ nm for the contour and
persistence lengths of the handle, $L^{\rm u}_{\rm ssRNA}=20.1$ nm
(34 bases) and $P_{\rm ssRNA}=1.0$ nm for the complete unfolded
RNA, $L^{\rm f}_{\rm ssRNA}=1.2$ nm (2 bases) for the folded RNA,
$\ln k^{\rm u}_0=-41.$ and $\ln k^{\rm f}_0=27.$ for the
logarithms of the unfolding and folding rates in the absence of
force, and $d_{\rm f}^\ddag=d_{\rm u}^\ddag=10$ nm for the
locations of transition state; all values are in the experimental
ranges~\cite{Wen,Woodside}. Additionally, we choose the cutoff
$k_{\rm max}\approx 4\times10^4$ $s^{-1}$, which is about ten
times bigger than the corner frequency in the
experiment~\cite{Wen}. We see that the simulations are
qualitatively consistent with the experimental
observation~\cite{Wen}. In the following we focus our attention on
the inference of the intrinsic kinetic parameters
from the time series obtained by simulation.\\

{\noindent\bf Bayesian parameter estimates.} Let
$\mathbf{x}=\left(x_0,\cdots,x_n\right)$ be a sequence of the
distances $x_l$ observed at equal separated time point $t_l$ at a
given constant force $F$ or $x_{\rm T}$ ($x_l=x_{\rm
T}-f_l/\varepsilon$ in the PM). According to Bayes' theorem, the
posterior distribution on the parameters $\theta=\left(\ln k^{\rm
u}_0,\ln k^{\rm f}_0,d_{\rm f}^\ddag,d_{\rm u}^\ddag\right)$ given
the observation $\mathbf{x}$ is
\begin{eqnarray}
\label{posteriordistribution}
P( \theta|\mathbf{x})\propto \eta(
\theta)L(\mathbf{x}|\theta ),
\end{eqnarray}
where $\eta(\theta)$ and $L(\mathbf{x}| \theta )$ are the prior
distribution on the parameters and likelihood function of
observing $\mathbf{x}$ given the parameters, respectively; the
reason we use the logarithms of the rates instead of themselves
will be seen soon.

The RNA is either folded or unfolded at any time. Because the
light tweezer experiment only records the distance between the
centers of the two beads, the folding/unfolding of single RNA is
virtually a hidden Markov process~\cite{Rabiner}. The likelihood
then is
\begin{eqnarray}
L({\mathbf x}|{\mathbf \theta }) = {\mathbf 1}^{T} \times
\prod\limits_{l=n}^{1} {\mathbf P}(x_l,t_l|x_{l-1},t_{l-1} )
\times {\mathbf P}_{0} (x_0).
\end{eqnarray}
The matrix element $\left[\mathbf P(x,\Delta t|y,0)\right]_{\rm i
j}$ (i,j=u,f) in the above equation represents the transition
probability of Eq.~\ref{diffusionreactonequations} with the
initial value $\delta_{\rm ij}{\text { }}\delta(x-y)$, and $\Delta
t=t_{l+1}-t_l$. We have also assumed the observation starting the
steady-state $\mathbf{P}_0 (x_0) = \left[ P^{\rm ss}_{\rm f}(x_0),
P^{\rm ss}_{\rm u}(x_0) \right ]^{T} $. We mentioned that
Eq.~\ref{diffusionreactonequations} usually does not have exact
time-dependent solutions. But in the real experiments the
relaxation time of the bead in the light tweezer is mostly shorter
than the measurement time and the relaxation time of the RNA
kinetics, namely, $\tau_{\rm b}\ll \Delta t, \tau_{\rm f-u}$. We
call such a case as rapid diffusion limiting ($D\to\infty$). Under
this limiting, we obtain
\begin{eqnarray}
\label{rapiddiffusionsolution} {\mathbf{P}}(x,\Delta t|y,0 )
\simeq {\mathbf{\Lambda }}(x ){\mathbf{Q}}(\Delta t),
\end{eqnarray}
where
\begin{eqnarray}
\label{positionprobability}
\mathbf{\Lambda}(x)={\rm
diag}\left[p^{\rm eq}_{\rm f}(x),p^{\rm eq}_{\rm u}(x)\right],
\end{eqnarray}
and
\begin{eqnarray}
\label{transitionprobability} {\mathbf{Q}}(\Delta t) = \left(
{\begin{array}{*{20}c}
   {\pi_{\rm f}  + \pi_{\rm u} e^{ - \Delta t\langle k\rangle} } & {\pi_{\rm f}
   \left (1 - e^{ - \Delta t\langle k\rangle} \right)}  \\
   {\pi_{\rm u} \left (1 - e^{ -\Delta  t\langle k\rangle} \right )} & {\pi_{\rm u}  +
   \pi_{\rm f} e^{ -\Delta  t\langle k\rangle} }  \\
 \end{array} } \right);
\end{eqnarray}
it is independent of the initial position of the bead $y$.  With
Eqs.~\ref{positionprobability} and \ref{transitionprobability},
the likelihood function can be calculated by the forward recursion
and ongoing scaling techniques~\cite{Rabiner}. On the other hand,
in order to have sufficient data to make reliable estimates of the
parameters, we use multiple observation sequences obtained at
different experimental control parameters, i.e., different
constant forces $F$ in the CFM or distances $x_{\rm T}$ in the PM.
The joint likelihood is simply a multiplication of
Eq.~\ref{posteriordistribution} at a certain force or distance.
Finally, we choose independent flat priors for the parameters in
$\theta$. Because we are treating the logarithms of the rates,
their flat priors are equivalent to the Jeffreys'
priors~\cite{Gelman} of the rates themselves.

\begin{figure}[htpb]
\begin{center}
\includegraphics[width=0.9\columnwidth]{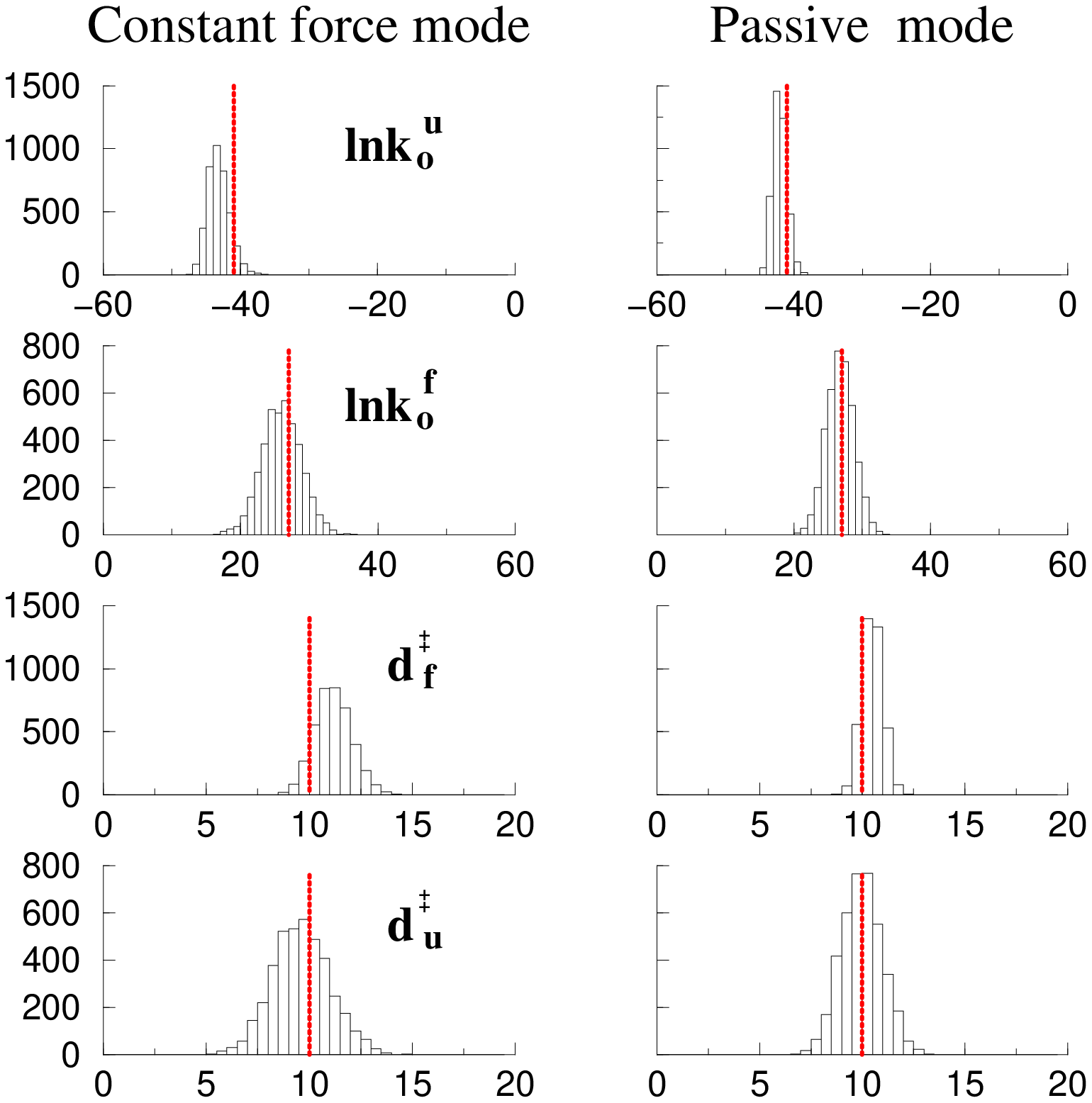}
\end{center}
\caption{} \label{figure3}
\end{figure}

Direct computation from $P( \theta|\mathbf{x})$ is infeasible. We
use standard Metropolis Monte Carlo algorithm~\cite{Gelman} to
sample from it. Fig.~\ref{figure3} illustrates the posterior
sampling distributions on the four parameters from two data sets
in the CFM and PM, respectively. Each data set is composed of five
time series simulated at five different control parameters: in the
CFM, $F$=11.7, 12.0, 12.3, 12.5, 13.0 pN, and in the PM, $x_{\rm
T}$=777, 780, 785 789, 795 nm. Their time interval and during time
are the same with those in Fig.~\ref{figure2}. Table~\ref{table1}
is the mean of these parameters inferred from ten data sets in the
two modes. We see that the means for the parameters obtained by
the Bayesian method are very accurate and the variances are fairly
small in the two modes.

It is interesting to evaluate the difference of the inferences of
the intrinsic kinetic parameters of the RNA by our Bayesian method
and by the traditional histogram fitting
method~\cite{Liphardt01,Wen,Woodside}. We see that the parameters
inferred by the latter method apparently deviate from the actual
values; see the third line in Table~\ref{table1}. In order to
exclude the possibility of inadequacy of the fitting data, we also
directly fit the mean folding/unfolding rate $\langle k^{\rm
i}\rangle_{\bar \rm i}$ (i=f,u) at different constant forces by
the Bell formula. The results (the second line in
Table~\ref{table1}) are consistent with those obtained by the
histogram fitting method. Therefore, the fluctuation of the force
applied on the RNA significantly modulates the force dependence of
the folding/unfolding rates in nonlinear way. Indeed, it is easily
seen from the ratio, $\ln \langle e^{\beta f_{\rm f}d_{\rm
f}^\ddag} \rangle_{\rm f}\left/\beta F\right.$, which is no longer
a constant even if $\langle f_{\rm f}(x)\rangle_{\rm f}=F$ in the
steady state.

In conclusion, we construct a coarse-grain physical model to
describe the kinetics of the forced folding/unfolding RNA in the
light tweezer done in the CFM and PM. This model has properly
taken into account of the RNA kinetics, the dynamics of the beads,
and the elasticity of handles and RNA molecule. Then based on an
analytic solution of the model, we apply Bayesian statistics to
infer the intrinsic kinetic parameters of the single RNA from the
time series of the distance or force. Our results show that, if
the fluctuation of the force is significant, which could be
induced by the Brownian motion of the bead in the light tweezer or
the structural transitions of the RNA, the traditional histogram
method would be problematic in inferring the intrinsic parameters.
Under this situation, the Bayesian method developed here would be
a better alternative.
\\

{\noindent}F.L. would like to thank Drs. Hu Chen and Jie Yan for
generously showing us their unpublished calculation about the
effective persistence of a sequence of heterogeneous WLCs. We also
appreciate Prof. Jian Wu for his great help in computation. This
work is funded by Tsinghua Basic Research Foundation.

\newpage
\begin{table} \caption{Means for the
intrinsic kinetic parameters inferred by our Bayesian method (BM)
in the CFM and PM and the traditional histogram fitting method
(HFM) in the CFM. Ten data sets are used here. As a comparison,
the parameters obtained by exact fitting (EF) the mean
folding/unfolding rates are also listed.} \label{table1}
\begin{center}
\begin{tabular}{ccccc}
\hline & $\ln k^+_0$& $\ln k^-_0$ & $d_{\rm f}^\ddag$  &  $d_{\rm u}^\ddag$  \\
\hline Actual value & -41. &  27. & 10. & 10.   \\
EF in CFM  &-16.9 & 24.9&6.5 & 7.3 \\
HFM in CFM  &$-15.7\pm 1.5$ &$23.0\pm 1.4$ & $6.2\pm 0.5$ & $6.7\pm 0.5$ \\
{\bf BM in CFM} & $-39.4\pm 4.2$ & $26.1\pm 3.3$ & $9.7\pm 1.1$ & $9.7\pm 1.6$ \\
{\bf BM in PM} & $-41.4\pm 1.5$ & $26.6\pm 1.4$ & $10.3\pm 0.7$ & $9.9\pm 0.7$ \\
\hline
\end{tabular}
\end{center}
\end{table}

\newpage
Fig captions:\\

Fig.1. (Color online.) Sketch of the forced folding/unfolding of a
RNA in a light tweezer. The RNA molecule is attached between the
two beads (larger red points) with two long DNA/RNA hybrid handles
(the black dash curves). In the constant force mode, a constant
force $F$ is exerted on the bead in the light tweezer. While in
the passive mode~\cite{Wen}, the distance between the centers of
the light tweezer and the bead held by micropipette is left
stationary, namely, $x_{\rm T}=x^{\rm tw}+x$ is a constant
($x=x^{\rm ds}_1+x^{\rm ss}+x^{\rm ds}_2$).  We do not include the
sizes of the beads in $x_{\rm T}$
for it does not matter to our discussion.\\

Fig.2. (Color online.) Time series of the distance $x$ at three
different constant forces in the CFM (left column) and of the
force exerted by the light tweezer at three different $x_{\rm T}$
in the PM (right column). The duration of them is 6 s and the time
interval is 1 ms.\\

Fig.3. (Color online.) Histograms of the posterior samples for one data set
generated by simulating Eq.~\ref{diffusionreactonequations} in the
CFM and PM, respectively. Each data set in the two modes is
composed of five time series obtained at five different control
parameters. The red vertical dashed lines in the panels represent
the actual parameters.

\newpage

\newpage

\newpage


\begin{thebibliography}{99}
\bibitem{Liphardt01}
J.B. Liphardt, {\it et al.}, \sci{292}{733}{2001}.

\bibitem{Woodside}
M.T. Woodside, {\it et al.}, \sci{314}{1001}{2006}.

\bibitem{Wen}
J.D. Wen, {\it et al.}, \bio{92}{2996}{2007}.

\bibitem{Manosas07}
M. Manosas {\it et al.}, \bio{92}{3010}{2007}.


\bibitem{Manosas05}
M. Manosas and F. Ritort, \bio{88}{3224}{2005}.


\bibitem{Marko}
J.F. Marko and E.D. Siggia, Macromolecules, {\bf 28}, 8759 (1995).

\bibitem {Bustamante}
C. Bustamante, J.F. Marko, E.D. Siggia, and S. Smith,
\sci{264}{1599}{1994}.

\bibitem{liuf1}
We do not need to model the handles and ssRNA chain independently,
because the effective persistent length of a sequence of connected
worm like chains (WLCs) can be calculated by the following
formula: $P_{\rm eff}=\left(\frac{L_1+\cdots+L_n}{L_1/\sqrt{P_1} +
\cdots+ L_n/\sqrt{P_n} }\right)^2$, where $L_i$ and $P_i$
(i=1,$\cdots$,n) are the contour lengthes and persistent lengthes
of the WLCs, respectively.  (Chen and Yan, personal
communications).

\bibitem{Bell}
G.I. Bell, \sci{200}{618}{1978}.

\bibitem{Dembo}
M. Dembo, {\it et al.}, Proc. R. Soc. Lond. B Biol. Sci. {\bf
234}, 55 (1988).

\bibitem{Evans97}
E. Evans and K. Ritchie, \bio{72}{1541}{1997}.

\bibitem{Shapiro}
B.E. Shapiro and H. Qian, Biophys. Chem. {\bf 67}, 211 (1997).

\bibitem{Dudkoprl}
O.K. Dudko, G. Hummer and A. Szabo \prl{96}{108101}{2006}.


\bibitem{liuf2}
It has been known that as force increases, the transtion state
location in any one-dimensional potential necessarily decreases
and thus weakens the dependence of the unfolding rate on the
force~\cite{Dudkoprl}. A cutoff of the rate here can be seen as
very rough approximation to this procedure. Of course, we can
replace this rate by other rate formulas having microscopic
foundation, which would be left in the analysis of real
single-molecule data.

\bibitem{liuf3}
Another extreme case, the slow diffusion limiting ($D\to 0$) is not of
interest in theory and experiment for the distance of $x$ of the
centers of the beads does not change with time.

\bibitem{Rabiner}
L.R. Rabiner, Proc. IEEE {\bf 77}, 257 (1989).

\bibitem{Gelman}
A. Gelman, J.B. Carlin, H.S. Stern, and D.B. Rubin, {\it Bayesian
Data Analysis} (Chapman and Hall, 1995).

\end{thebibliography}
\end{document}